\documentclass{PoS}
\usepackage{subfigure}

\newcommand{\BR}{\ensuremath{\mathcal B}}

\title{Rare decay searches at CDF}

\ShortTitle{Rare decay searches at CDF}

\author{\speaker{Paolo Maestro}\\ On behalf of the CDF collaboration\\
        Department of Physics, University of Siena and INFN, via Roma 56, 53100 Siena, Italy\\
        E-mail: \email{paolo.maestro@pi.infn.it}}

\abstract{
In the last decade the CDF  experiment at the Tevatron clearly demonstrated
that it is possible 
to study extensively heavy flavour physics in hadron collisions and  achieve remarkable results, competitive 
and complementary to $B$-factories.
In this paper we report on the indirect searches for physics beyond the standard model via measurements
of rare $b$-hadron decays. The final limits, based on the analysis of the full CDF
data set, on the branching fraction of the $B^0_{(s)}$  decay  into a pair of muons are presented and discussed.
Moreover we review the latest measurements, with 6.8 fb$^{-1}$ of collected data, 
of the total and differential branching fractions and angular observables 
of rare  $b$-hadron decays  proceeding via the flavour-changing neutral-current 
process $b \rightarrow s \mu^+ \mu^-$.
}

\FullConference{The XIth International Conference on Heavy Quarks and Leptons,\\
		June 11-15, 2012\\
		Prague, Czech Republic}

\begin{document}

\section{Introduction}
Rare flavour-changing neutral-current (FCNC) $B^0_{(s)} \rightarrow \mu^+ \mu^-$  and $b \rightarrow s \mu^+ \mu^-$ decays
are considered among the most promising  probes
of the standard model (SM) and its extensions.
The precise measurement of several observables (total and differential
branching ratios, angular distributions of the decay products) of these decays
might provide interesting clues for new physics (NP) phenomena,
if any sizable deviation  from the SM predictions is observed.
In this paper we review the current status of these indirect searches at the Collider 
Detector at Fermilab (CDF II), which
reached a sensitivity very close to SM predictions after a 
decade of Tevatron leadership in the exploration of $B^0_s$ dynamics.
The Tevatron $p \bar{p}$ collider, whose operations ended in October 2011 after 20 years of operation, 
provided excellent opportunities to study $B$ physics. % by rich $b$ quark production and its hadronization.
CDF II is a multipurpose detector, consisting of a central charged particle tracking system, 
surrounded by calorimeters and muon chambers. It collected a final data set corresponding
to about 10 fb$^{-1}$ of integrated luminosity. 

In the search for rare $B$ decays,
the  experimental challenge is to reject a huge background while keeping the signal efficiency high.
A dedicated dimuon trigger has been used to select 
events with a pair of muons in the final state in the pseudorapidity region $|\eta|<$1.1.
%%The relevant CDF  figures  of merit for this analysis are 
In the reconstruction and analysis of $B$ hadron decays, CDF takes advantage of
%Moreover, the 
an excellent transverse momentum resolution, 
%\begin{equation}
$\frac{\sigma_{p_T}}{p_T} = 0.07\%\, p_T$ (GeV/c),
%\end{equation}
which implies a resolution on the invariant
dimuon mass of 24 MeV in the $B^0_{(s)} \rightarrow \mu^+ \mu^-$ decay, 
a vertex resolution  of about 30 $\mu$m in the transverse plane, and 
particle identification (PID) capability, 
based on multiple measurements of the ionization per unit of path length ($dE/dx$) in the drift chambers.  
%are the relevant CDF figures of merit exploited in these analyses to select clean samples of rare $b$-hadron decays.
%%%%%%%%%%%%%%%%%%%%%%%%%%%%%%%%%%%%%%%%%%%%%%%%%%%%%%%%%%%%%%%%%%%%%%%%%%%%%%
\section{Search for $B^0_{(s)} \rightarrow \mu^{+} \mu^{-}$}
$B_{(s)}^0 \rightarrow \mu^{+} \mu^{-}$ decays are mediated by FCNC and thus forbidden at first order in the SM. 
Moreover they are further suppressed by helicity factors $(m_{\mu}/m_B)^2$ in the final dimuon state.
They can only occur at second order through penguin and box diagrams. 
The SM predicts very low rates for these processes: $\BR(B_{s}^0 \rightarrow \mu^{+} \mu^{-}) = \left(3.2 \pm 0.2\right)\times 10^{-9}$ 
and  $\BR(B^0 \rightarrow \mu^{+} \mu^{-}) = \left(1.0 \pm 0.1\right)\times 10^{-10}$ \cite{SM1,SM2}.
However, a wide variety of beyond the standard model (BSM) theories predict enhancement of their branching ratios
 by several order of magnitudes, 
making these decays one of the most sensitive probes in indirect searches for NP. 

In 2011, CDF observed an intriguing $\sim$2.5$\sigma$ excess over background in
$B_{s}^0 \rightarrow \mu^{+} \mu^{-}$ using  7 fb$^{-1}$ of data \cite{CDF}.
Though it was compatible with other experimental results (LHCb \cite{LHCb}, CMS \cite{CMS})
and the SM prediction, it could be interpreted as the first indication of a signal
and allowed CDF to set a two-sided bound on the rate 
$\BR(B_{s}^0 \rightarrow \mu^{+} \mu^{-}) = \left(1.8^{+1.1}_{-0.9}\right)\times 10^{-8}$ .
To further investigate the nature of the excess, we repeated the analysis unchanged using the whole
%Since this result was widely debated, it became mandatory to analyze with 
%the  procedure unchanged the
 Run II data set, corresponding to 9.7 fb$^{-1}$ of integrated
luminosity, about 30\% more data with respect to the 2011 analysis. Here we report on 
the final results of this search \cite{CDF2}.
%%%%%%%%%%%%%%%%%%%%%%%%%%%%%%%%%%%%%%%%%%%%%%%%%%%%%%%%%%%%%%%%%%%%%%%%%%%%%%
\begin{figure}
  \centering
\subfigure[]{\label{fig:bmumu1}
  \includegraphics[width=.85\textwidth]{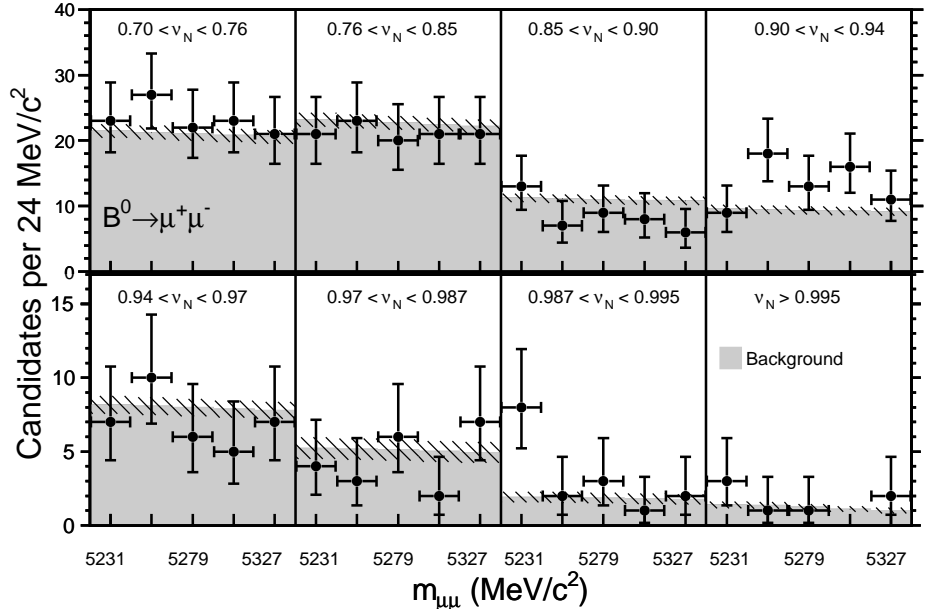}
}
\subfigure[]{\label{fig:bmumu2}
  \includegraphics[width=.85\textwidth]{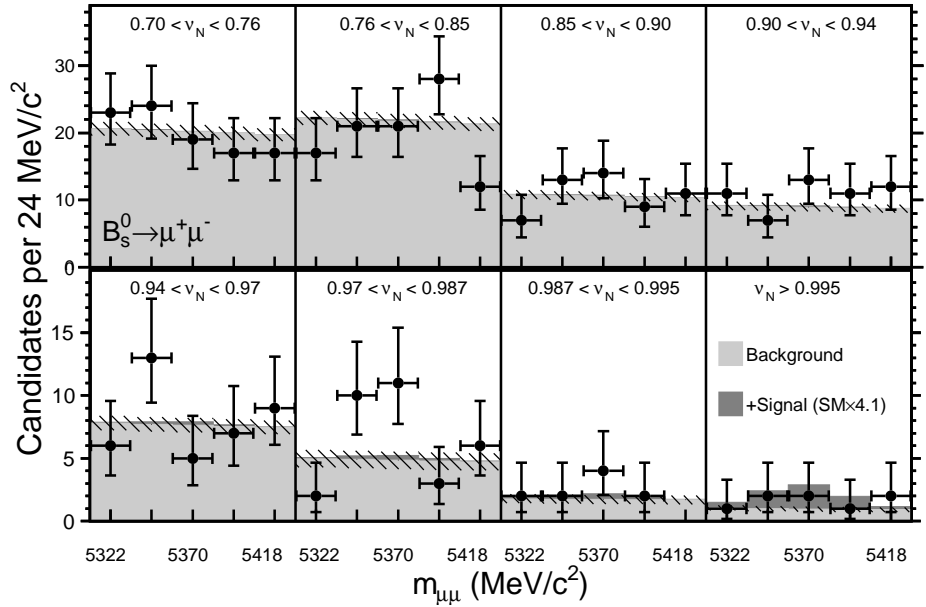}
}
\caption{
Dimuon mass distributions for the (a) $B^0 \rightarrow \mu^+ \mu^-$ and (b) 
$B_s^0 \rightarrow \mu^+ \mu^-$ signal region in the eight NN bins. 
The observed data (points) are compared to the total background expectation (light gray
histogram). The hatched region is the total uncertainty on the background expectation.
In (b) the dark gray histogram represents the SM signal expectation enhanced by a factor 4.1. 
}
\label{fig:bmumu}
\end{figure}
%%%%%%%%%%%%%%%%%%%%%%%%%%%%%%%%%%%%%%%%%%%%%%%%%%%%%%%%%%%%%%%%%%%%%%%%%%%%%%

The baseline selection  requires high quality muon candidates with opposite charge,
 transverse momentum $p_T >$ 2 GeV/c, and a dimuon invariant mass $m_{\mu\mu}$ in the range 4.669-5.969 GeV/c$^2$. 
The muon pairs are constrained to originate from a common, well-measured reconstructed decay point. 
A likelihood-based muon identification method, 
%based on information from muon systems and $dE/dx$ measured in the central tracker,
is used to suppress contributions 
from hadrons misidentified as muons. 
The branching ratios of $B_{(s)}^0 \rightarrow \mu^{+} \mu^{-}$ are measured
by normalizing to a sample of 40225$\pm$ 267
 $B^+ \rightarrow J/\psi (\rightarrow \mu^+ \mu^-)\,K^+$ candidates, 
selected with the same baseline requirements.
A Neural Network (NN) classifier is used to improve signal over background  separation. 
Fourteen variables are used to construct the NN discriminant that ranges between 0 and 1. 
The six most discriminating variables 
include the 3D opening angle between the dimuon momentum and the displacement vector between the
primary and secondary vertex; 
the isolation $I$ \footnote{$I = |\vec{p}_T^{\mu\mu}|/(\sum_i p_T^i + |\vec{p}_T^{\mu\mu}|)$, where $\vec{p}^{\mu\mu}$ is the momentum
of the dimuon pair; the sum is over all tracks with $\sqrt{ (\Delta\phi)^2 + (\Delta\eta)^2} \le 1$; $\Delta\eta$ and $\Delta\phi$ are the relative azimuthal angle and pseudorapidity of track $i$ with respect to $\vec{p}^{\mu\mu}$.}
of the candidate $B^0_{s}$ ;
%that is the number of tracks found within a cone defined by the muons momenta; 
the muon and $B^0_{(s)}$ impact parameters; the $B^0_{(s)}$ decay length significance; the vertex-fit $\chi^2$.
The NN from the 2011 analysis was used with the same training.
%The NN was trained with Monte Carlo (MC) signal and background events sampled from the mass sidebands. 
The final search region in the dimuon invariant mass has a half width of about 60 MeV corresponding to 2.5 times the dimuon 
mass resolution. 
The NN was validated with signal and background events. %the normalization mode and control regions. 
Careful checks for possible mass-biases  of the NN output and overtraining show no anomalies. 

Extensive and detailed background estimates and checks have been performed.
Background is due to  both combinatorial and peaking contributions in the signal region. 
Combinatorial background is estimated by fitting the sidebands  to linear functions, % with common slope, 
after blinding the signal region in the dimuon mass distribution.
%Alternative polynomial fit have been done to estimate uncertainty due to the function shape. 
The peaking background is due to $B \rightarrow h^+ h^{'-}$ decays where the hadrons 
($h,h'$ stand for $\pi$ or $K$) are misidentified as  muons. 
This has been estimated from both MC and data.  
The misidentification probability is parametrized as a function of the track transverse momentum
 using $D^*$-tagged $D^0 \rightarrow \pi^+ K^-$ events. 
It turned out that the peaking background is  about 10\% of the combinatorial background 
in $B_s^0 \rightarrow \mu^+ \mu^-$ and about 50\% of the total background in 
the $B^0 \rightarrow \mu^+ \mu^-$ channel.
Background estimates have been cross-checked using independent background-dominated control samples, 
in which the muons have the same measured charge or the reconstructed
dimuon candidate lifetime is negative. 
No significant discrepancies between the expected and observed number of events in the control samples have been found.\\
The data are divided into 8 bins of the NN discriminant to exploit the improved background suppression at high NN values, 
and five bins of mass in the search region.
In the $B^0$ search region  data are consistent with the background 
prediction (Fig.~\ref{fig:bmumu1}) and yield the limit of
 $\mathcal{B}(B^0 \rightarrow \mu^+ \mu^-) < 3.8 (4.6)\times 10^{-9}$ at 90\% (95\%) C.L.. 
The significance of the background-only hypothesis expressed as a p-value, estimated from an 
ensemble of background-only pseudo-experiments, is 41\%.

In the $B^0_s$ search region, a moderate excess in the highest NN bins ($>$0.97)
is observed (Fig.~\ref{fig:bmumu2}).
%the background prediction in bins with NN > 0.97. The bottom part of Fig. 3
%contains the detailed breakdown of the 
%expected background and actual observation for $B_s^0 \rightarrow \mu^+ \mu^-$
%the individual NN and mass bins for the search. 
The p-value for the
background-only hypothesis is 0.94\%. We also
produce an ensemble of simulated experiments that includes
a  $B^0_s \rightarrow \mu^+ \mu^-$ contribution at the expected SM
branching fraction which yields a p-value of 7.1\%.
%Expected SM yield of Bs ranges from 0.05 in lowest NN bin to 1.0 in highest. For Bd SM yield ~30 times smaller.
With respect to the previous CDF result with 7 fb$^{-1}$ of data \cite{CDF}, the excess in the third significant NN bin
(0.97-0.987)  softened, as expected for a statistical fluctuation.
Though the 2011 hint of signal is not reinforced by the new data,
it is still present and  remains $>$2$\sigma$ significant over background.
Assuming the observed
excess in the $B^0_s$ region is due to signal, CDF finds 
$\mathcal{B}(B_s^0 \rightarrow \mu^+ \mu^-) = \left(1.3^{+0.9}_{-0.7}\right)\times 10^{-8}$, 
which is still compatible with both
 the SM expectation and the latest  constraints from LHC experiments \cite{LHCbmumu, LHCb2}.

%%%%%%%%%%%%%%%%%%%%%%%%%%%%%%%%%%%%%%%%%%%%%%%%%%%%%%%%%%%%%%%%%%%%%%%%%%%%%%
\begin{figure}[!h]
  \centering
\subfigure[]{%\label{fig:bsmumu1}
  \includegraphics[width=4.5cm]{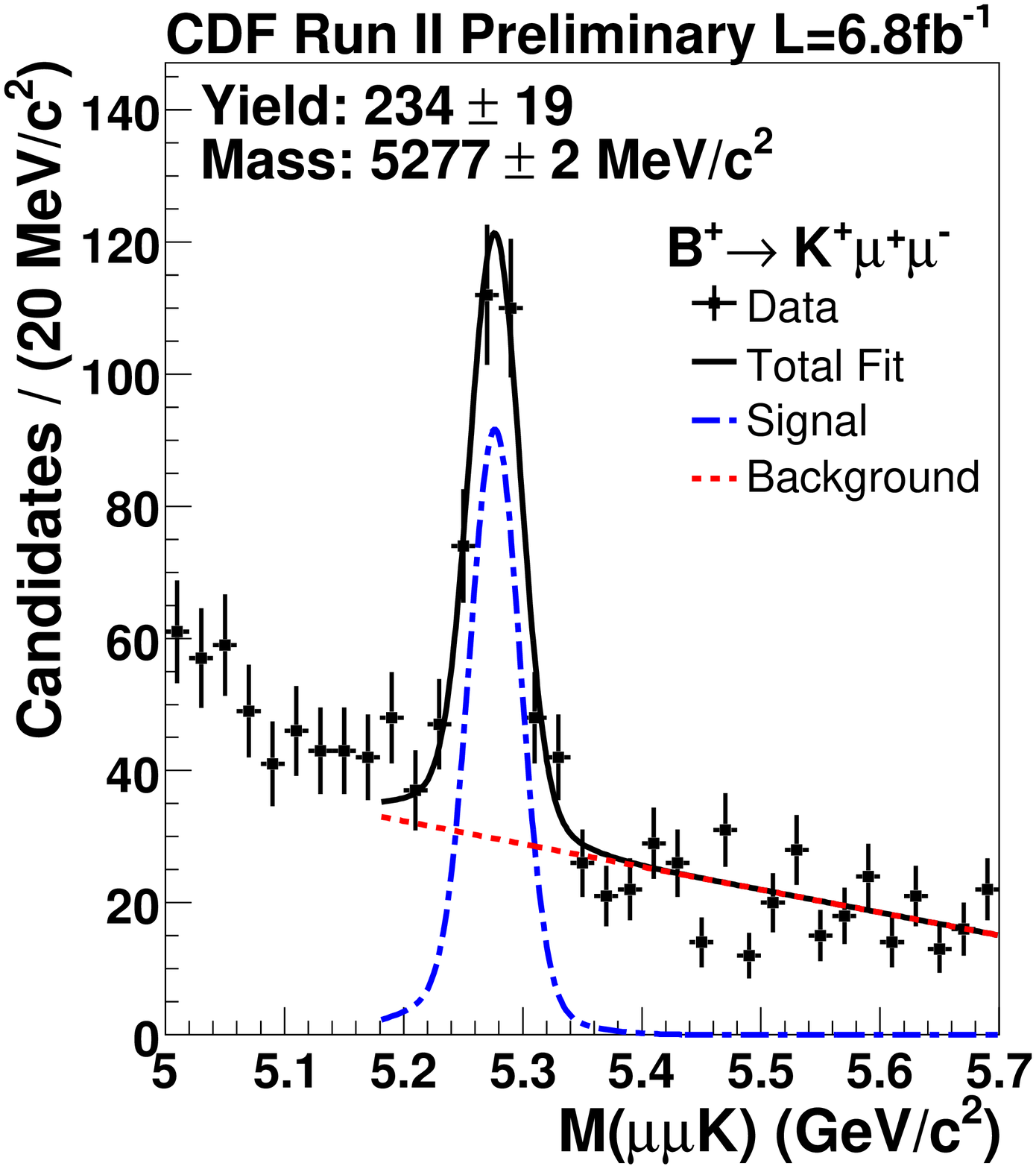}
}
\subfigure[]{%\label{fig:bsmumu2}
  \includegraphics[width=4.5cm]{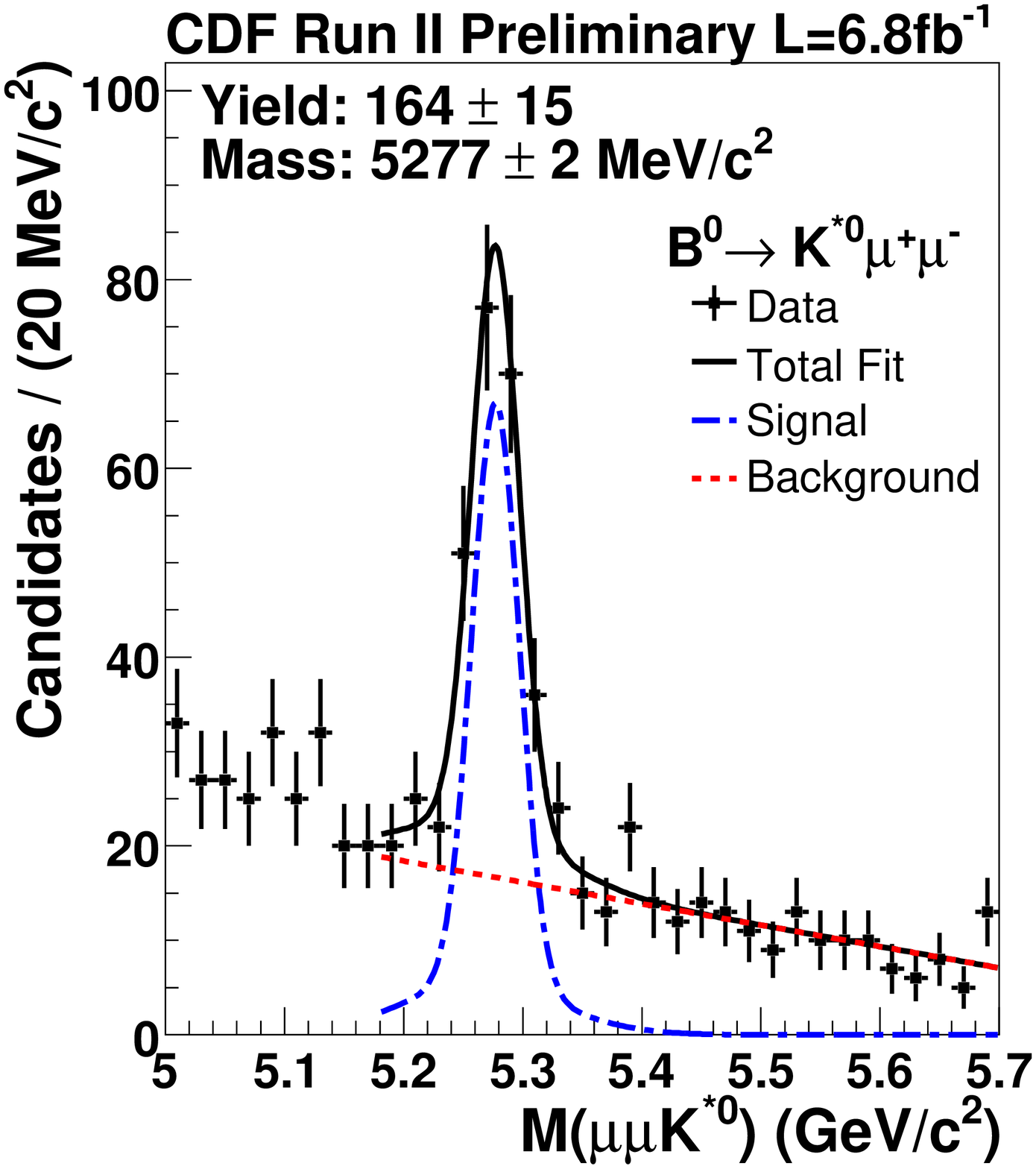}
}
\subfigure[]{%\label{fig:bsmumu3}
  \includegraphics[width=4.5cm]{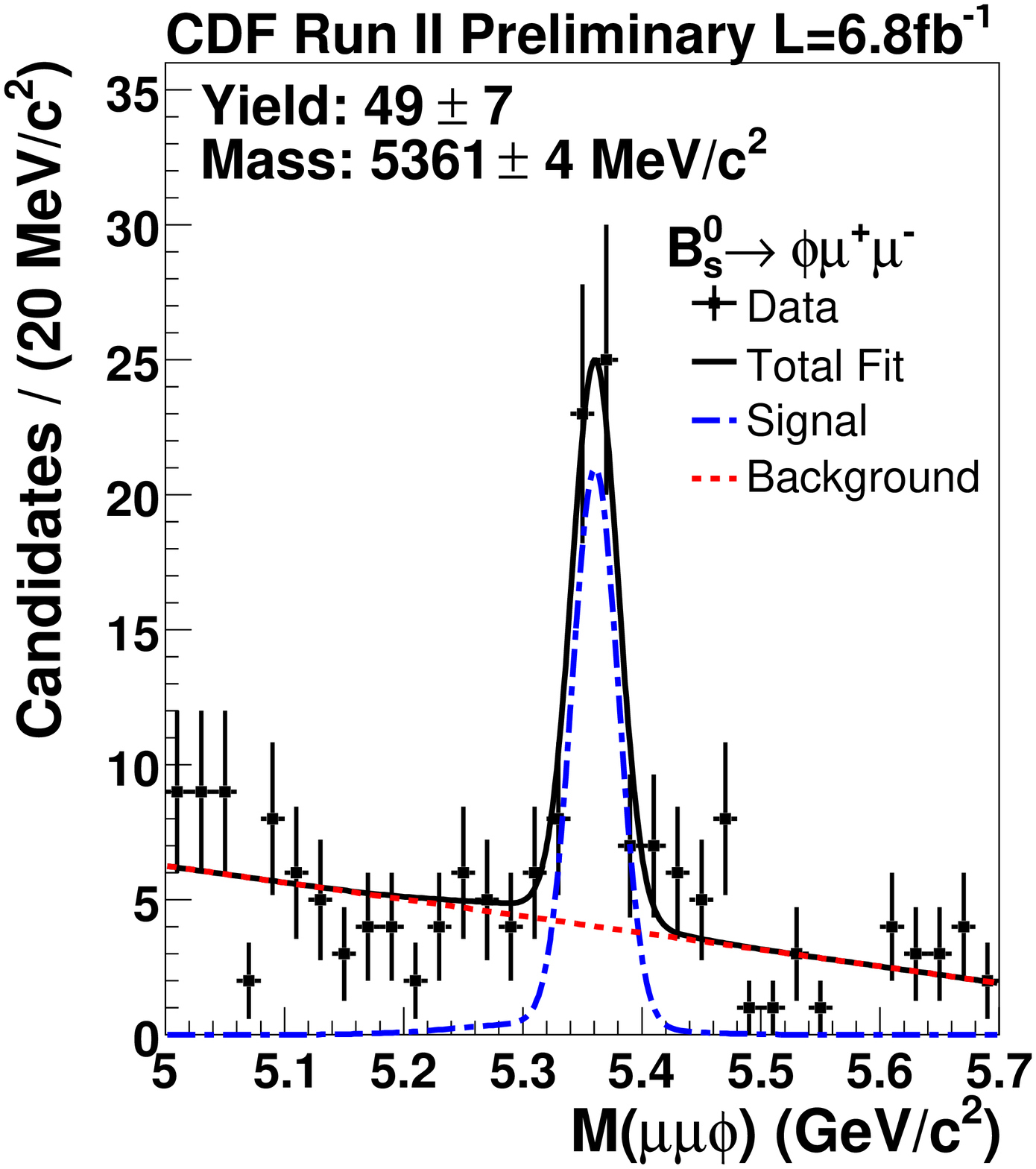}
}
\subfigure[]{%\label{fig:bsmumu4}
  \includegraphics[width=4.5cm]{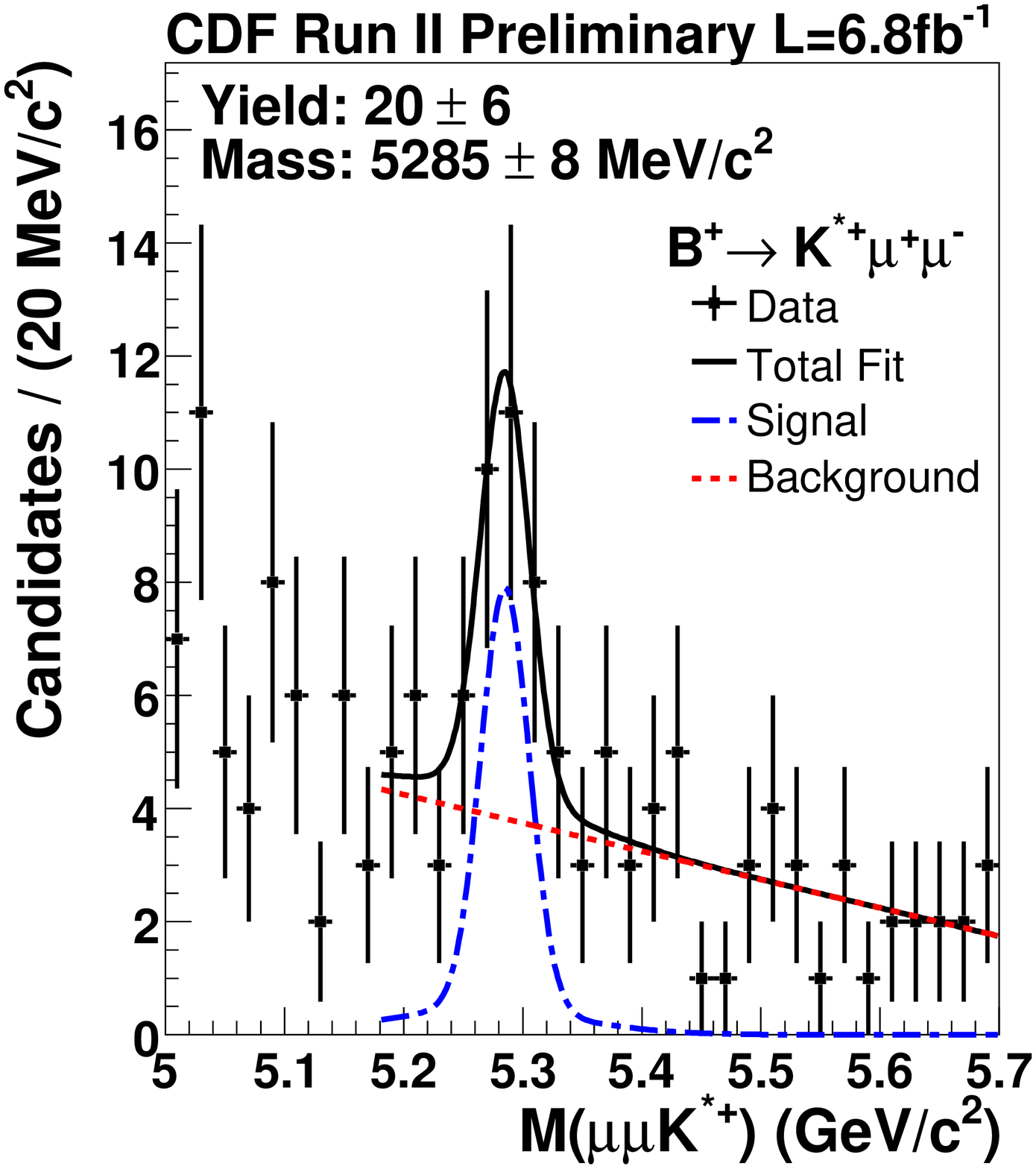}
}
\subfigure[]{%\label{fig:bsmumu5}
  \includegraphics[width=4.5cm]{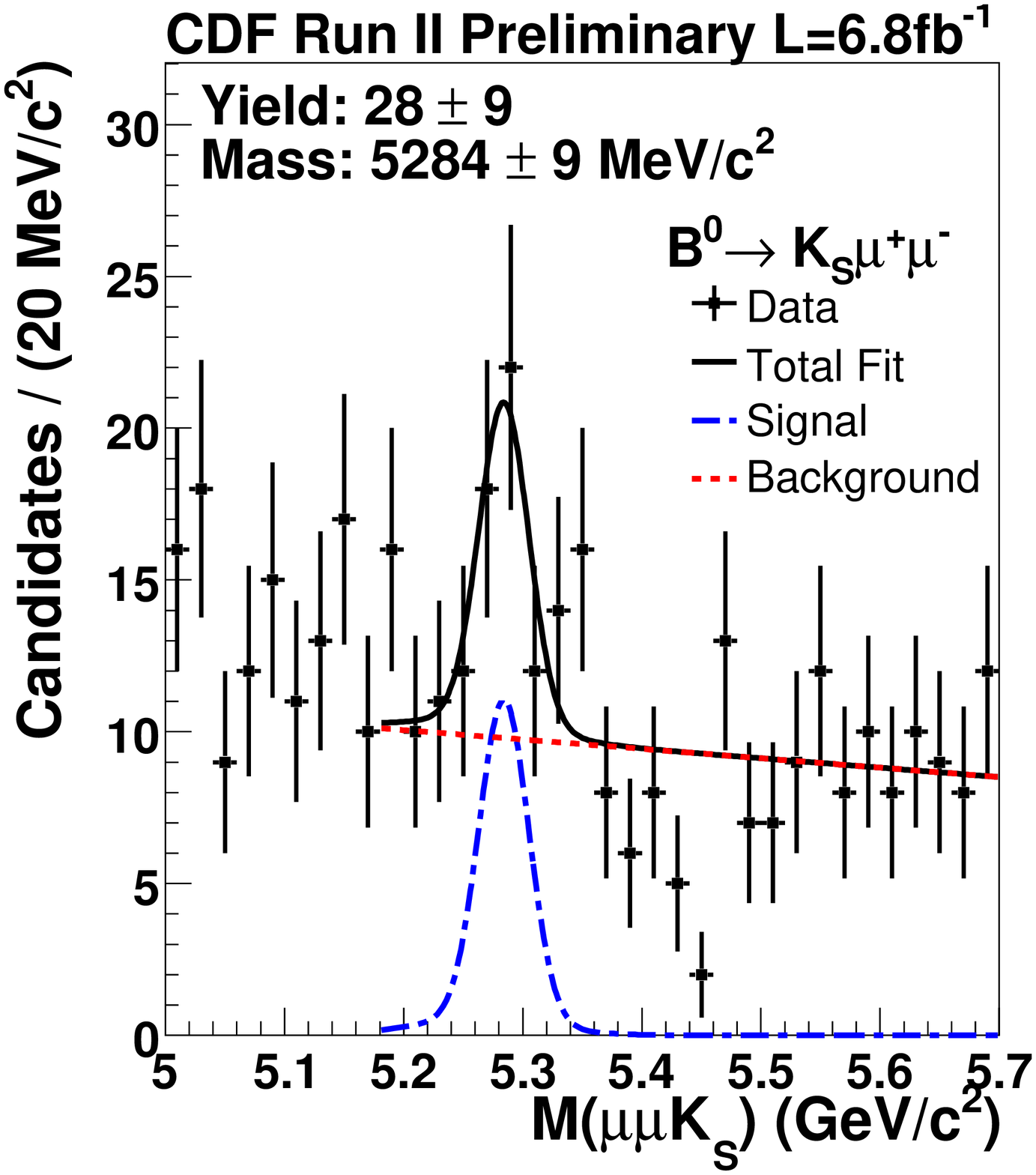}
}
\subfigure[]{%\label{fig:bsmumu6}
  \includegraphics[width=4.5cm]{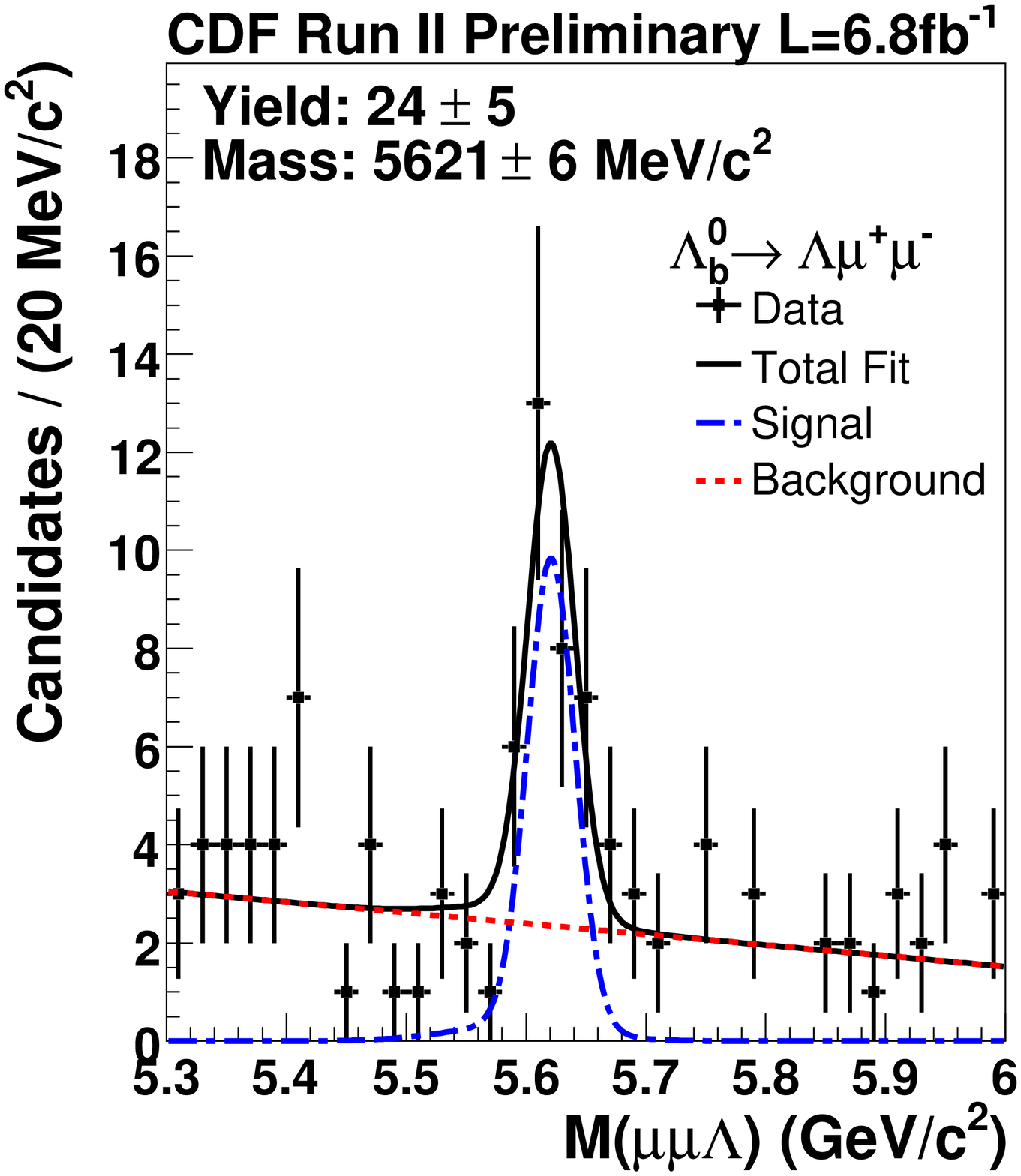}
}
\caption{Invariant mass of
(a) $B^+ \rightarrow K^+ \mu^+ \mu^- $, 
(b) $B^0 \rightarrow K^{*0} \mu^+ \mu^-$,
(c) $B^0_s \rightarrow \phi \mu^+ \mu^-$,
(d) $B^+ \rightarrow K^{*+} \mu^+ \mu^-$,
(e) $B^0 \rightarrow K_S \mu^+ \mu^-$, 
(f) $\Lambda_b^0 \rightarrow \Lambda \mu^+ \mu^-$ with fit results overlaid. }
\label{fig:bsmumuyield}
\end{figure}
%%%%%%%%%%%%%%%%%%%%%%%%%%%%%%%%%%%%%%%%%%%%%%%%%%%%%%%%%%%%%%%%%%%%%%%%%%%%%%
%%%%%%%%%%%%%%%%%%%%%%%%%%%%%%%%%%%%%%%%%%%%%%%%%%%%%%%%%%%%%%%%%%%%%%%%%%%%%%
\section{$b \rightarrow  s \mu^{+} \mu^{-}$  decays}
Rare decays of bottom hadrons mediated by the FCNC process $b \rightarrow s \mu^+ \mu^-$ 
are suppressed at tree level in the SM and must
occur through higher-order loop amplitudes. Their expected branching ratios are of the order of 10$^{-6}$.
Because of their clean experimental signature and the reliable theoretical predictions for their rates, 
these are excellent channels for NP searches.

%%%%%%%%%%%%%%%%%%%%%%%%%%%%%%%%%%%%%%%%%%%%%%%%%%%%%%%%%%%%%%%%%%%%%%%%%%%%%%%%
\begin{table}[h]
\begin{center} 
  \begin{tabular}{l l}
    \hline\hline Decay mode & $\mathcal{B}(10^{-6})$  \\
    \hline
    $B^+ \rightarrow K^+ \mu^+ \mu^- $ & $0.46 \pm 0.04 \pm 0.02 $\\
    $B^0 \rightarrow K^{*0} \mu^+ \mu^-$ &  $1.02 \pm 0.10 \pm 0.06$\\
    $B^0_s \rightarrow \phi \mu^+ \mu^-$ &  $1.47 \pm 0.24 \pm 0.46$\\
    $B^+ \rightarrow K^{*+} \mu^+ \mu^-$ & $0.95 \pm 0.32 \pm 0.08$\\
    $B^0 \rightarrow K_S \mu^+ \mu^-$ & $0.32 \pm 0.10 \pm 0.02$\\
    $\Lambda_b^0 \rightarrow \Lambda \mu^+ \mu^-$ & $1.73 \pm 0.42 \pm 0.55$\\
    \hline\hline
  \end{tabular}
\caption{Branching ratio of the $H_b\rightarrow h \mu^+ \mu^-$ decays measured by CDF.
The first quoted uncertainty is statistical, the second is systematic.}
  \label{tabBR}
\end{center}
\end{table}
%%%%%%%%%%%%%%%%%%%%%%%%%%%%%%%%%%%%%%%%%%%%%%%%%%%%%%%%%%%%%%%%%%%%%%%%%%%%%%
%Various physics backgrounds are reduced using mass vetoes. 
%Dimuon invariant mass is required to be inconsistent with decaying from mesons and B into charm decays
CDF has studied the  FCNC $H_b\rightarrow h \mu^+ \mu^-$ decays (where $H_b$ and $h$ indicate hadrons 
containing a $b$ and $s$ quark, respectively) listed in Table~\ref{tabBR}, 
using 6.8 fb$^{-1}$ of data collected with the dimuon trigger \cite{CDFbsmumuBR}. 
Candidates for each decay have been selected by standard kinematics cuts and a NN optimized for best sensitivity.
Signal yields are obtained by an unbinned maximum log-likelihood fit to the $b$-hadron mass distributions 
(Fig.~\ref{fig:bsmumuyield}), modelling 
the signal peak with a Gaussian, and the combinatorial background with a linear function.
To cancel dominant systematic uncertainties, the branching ratio of each rare decay  $H_b\rightarrow h \mu^+ \mu^-$ 
is measured relative to the corresponding resonant channel $H_b\rightarrow J/\psi\, h$,
used as a normalization and a cross-check of the whole analysis.
The results of the total branching ratios are reported in Table~\ref{tabBR} and include
the first observation of the baryonic FCNC decay $\Lambda_b^0 \rightarrow \Lambda \mu^+ \mu^-$, 
and the first measurement  of the $B^+ \rightarrow K^{*+} \mu^+ \mu^-$ and  $B^0 \rightarrow K_S\,\mu^+ \mu^-$ decays
at a hadron collider.
%%%%%%%%%%%%%%%%%%%%%%%%%%%%%%%%%%%%%%%%%%%%%%%%%%%%%%%%%%%%%%%%%%%%%%%%%%%%%%
\begin{figure}[h] 
  \centering
\subfigure[]{%\label{fig:bsmumu1}
  \includegraphics[width=4.5cm]{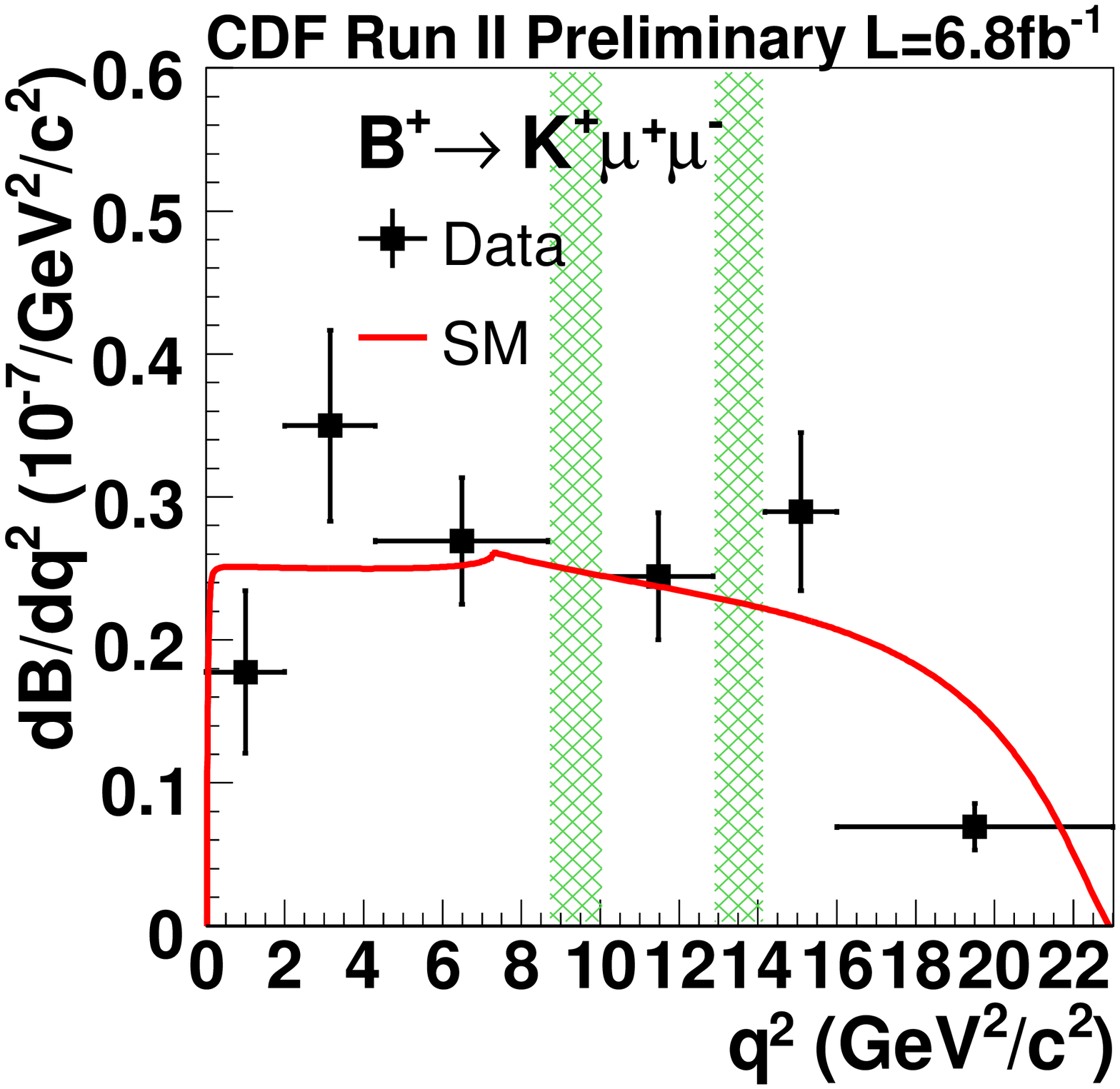}
}
\subfigure[]{%\label{fig:bsmumu2}
  \includegraphics[width=4.5cm]{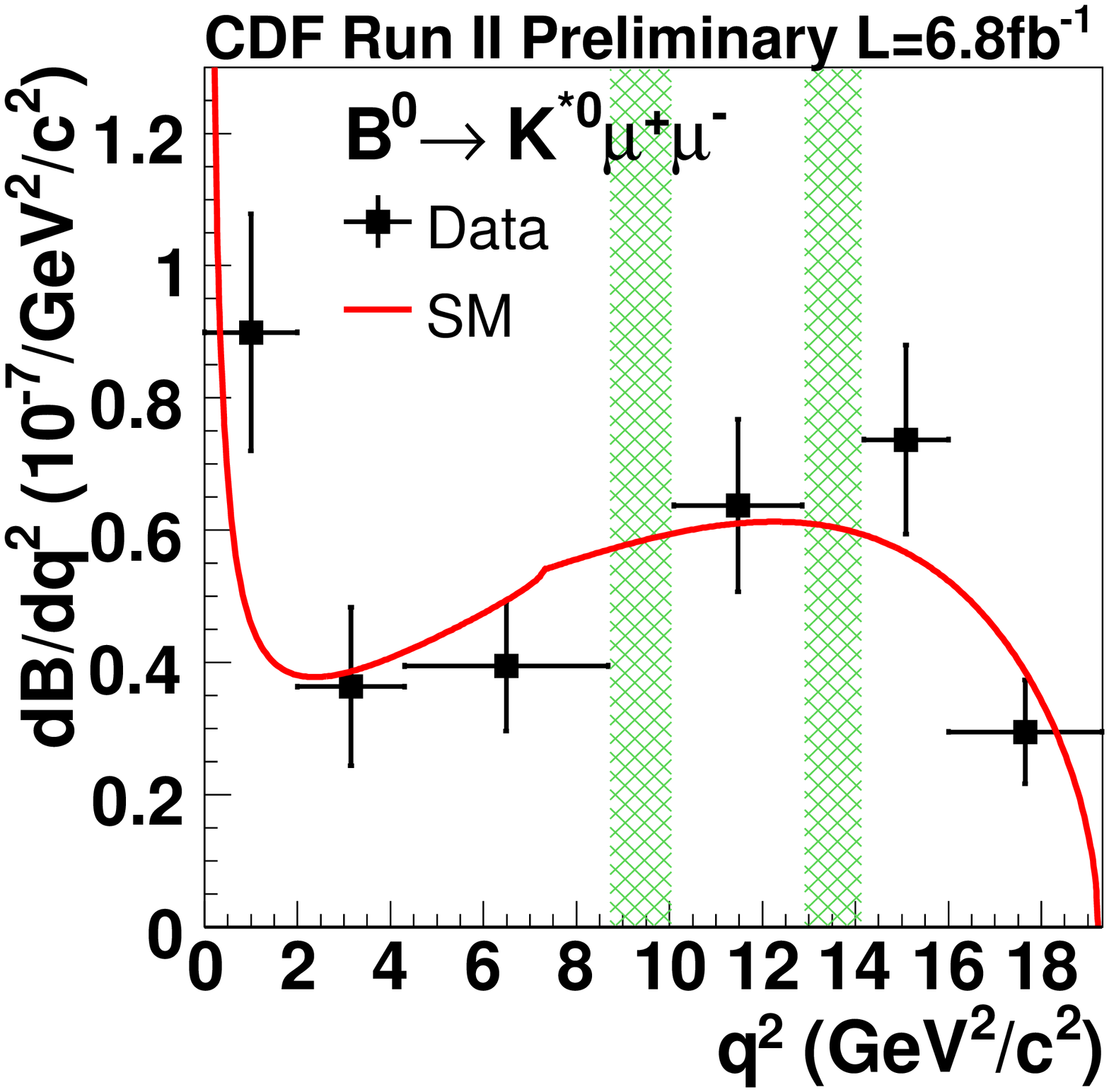}
}
\subfigure[]{%\label{fig:bsmumu3}
  \includegraphics[width=4.5cm]{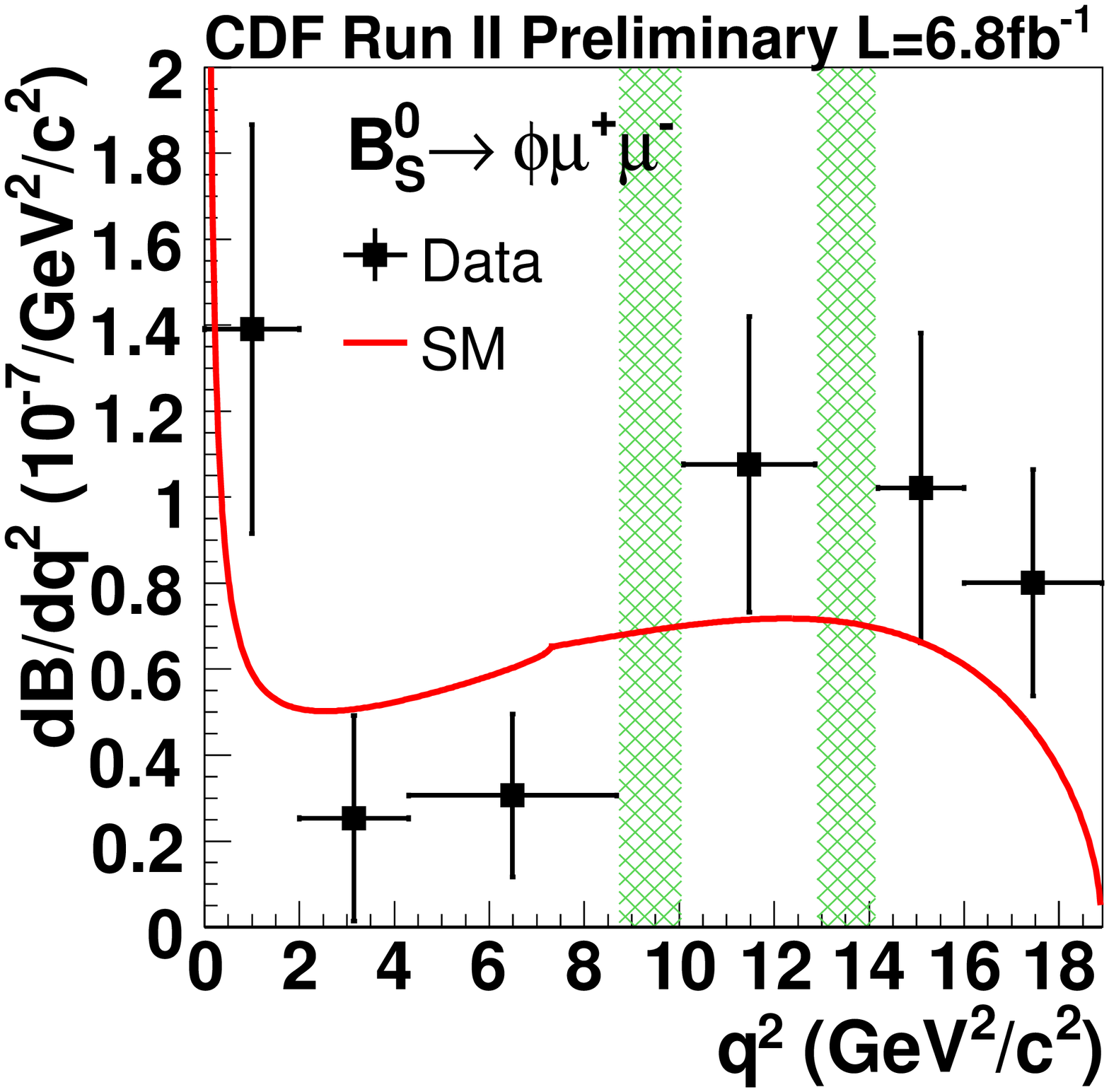}
}
\subfigure[]{%\label{fig:bsmumu4}
  \includegraphics[width=4.5cm]{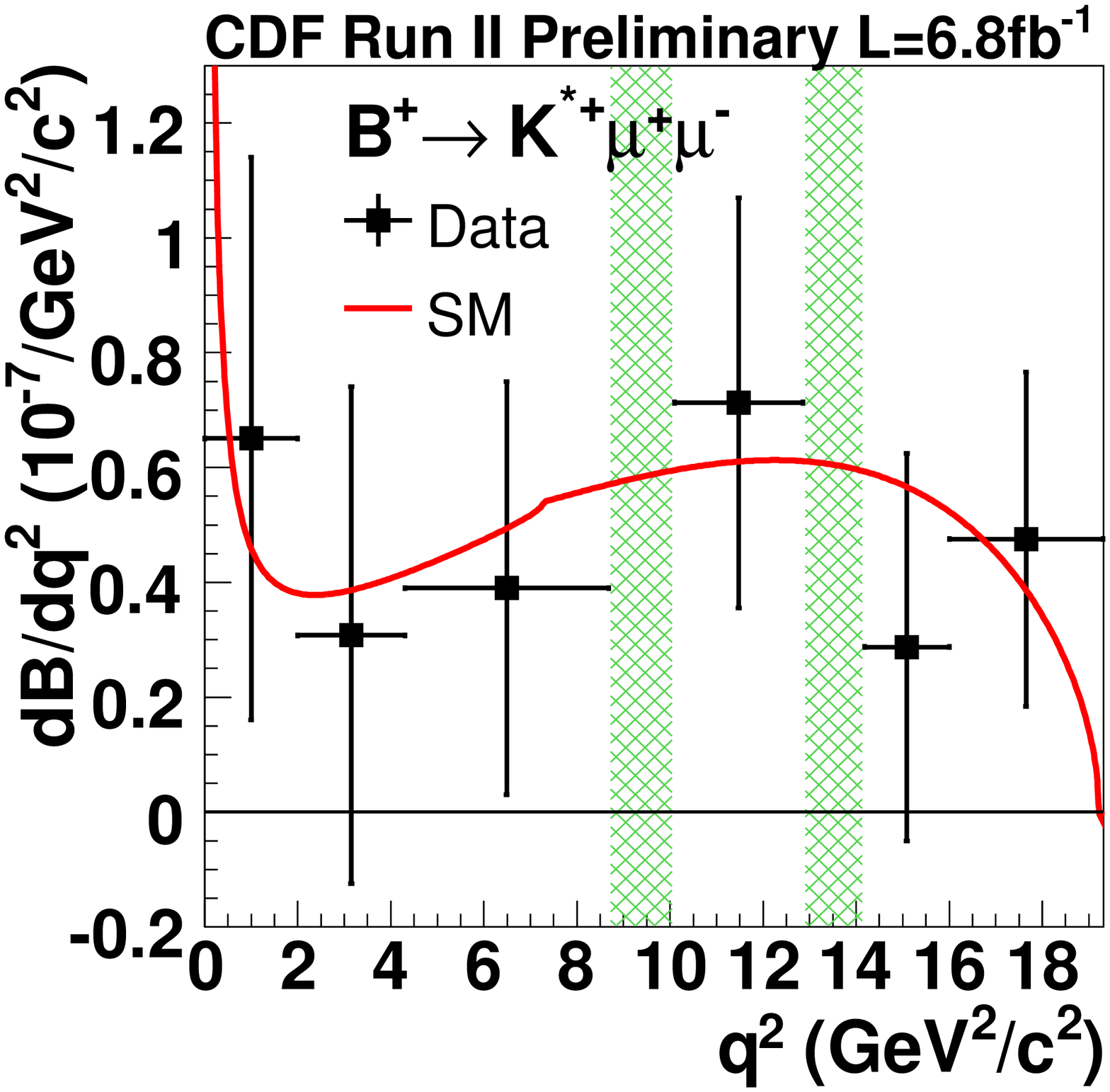}
}
\subfigure[]{%\label{fig:bsmumu5}
  \includegraphics[width=4.5cm]{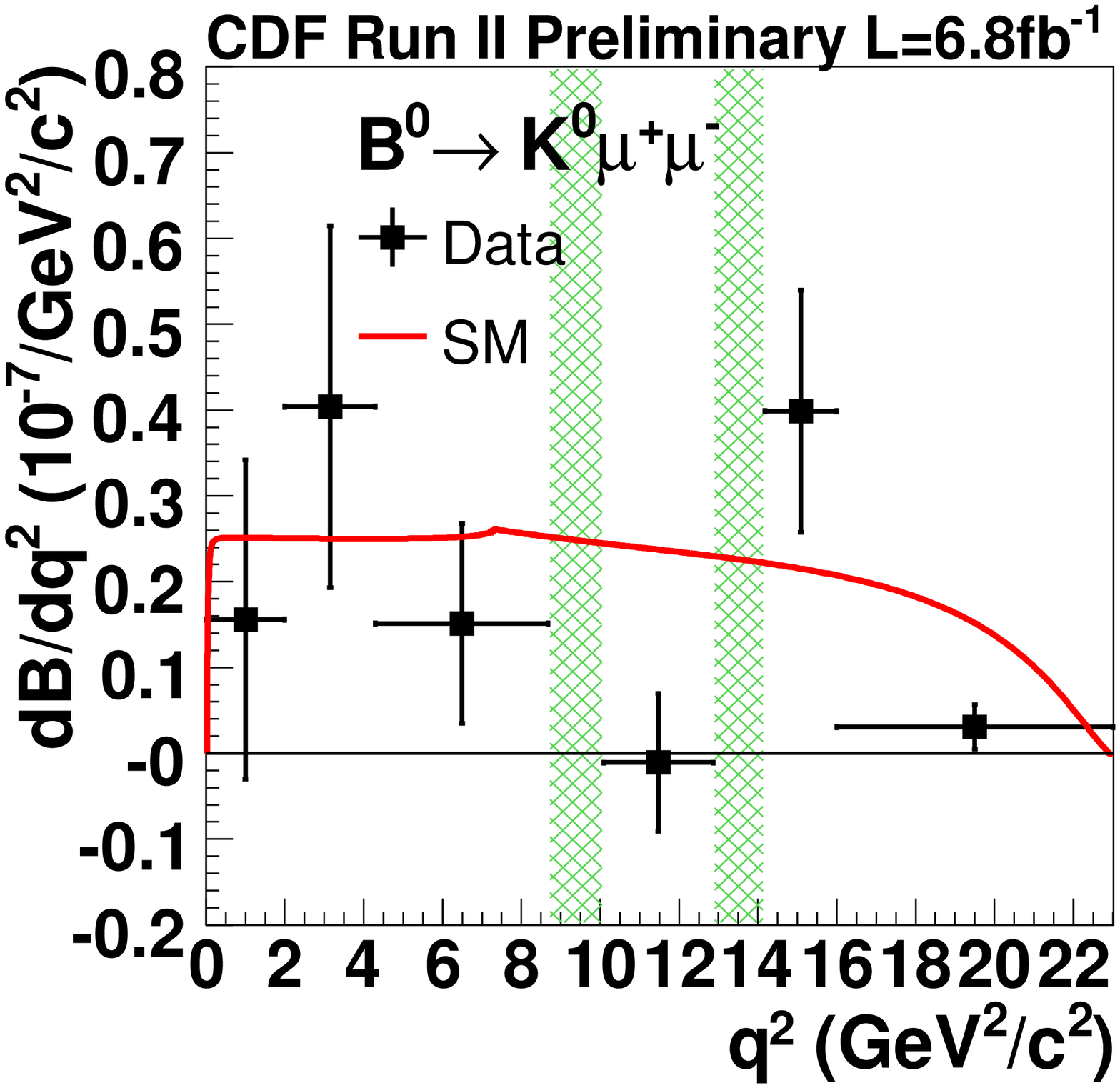}
}
\subfigure[]{%\label{fig:bsmumu6}
  \includegraphics[width=4.5cm]{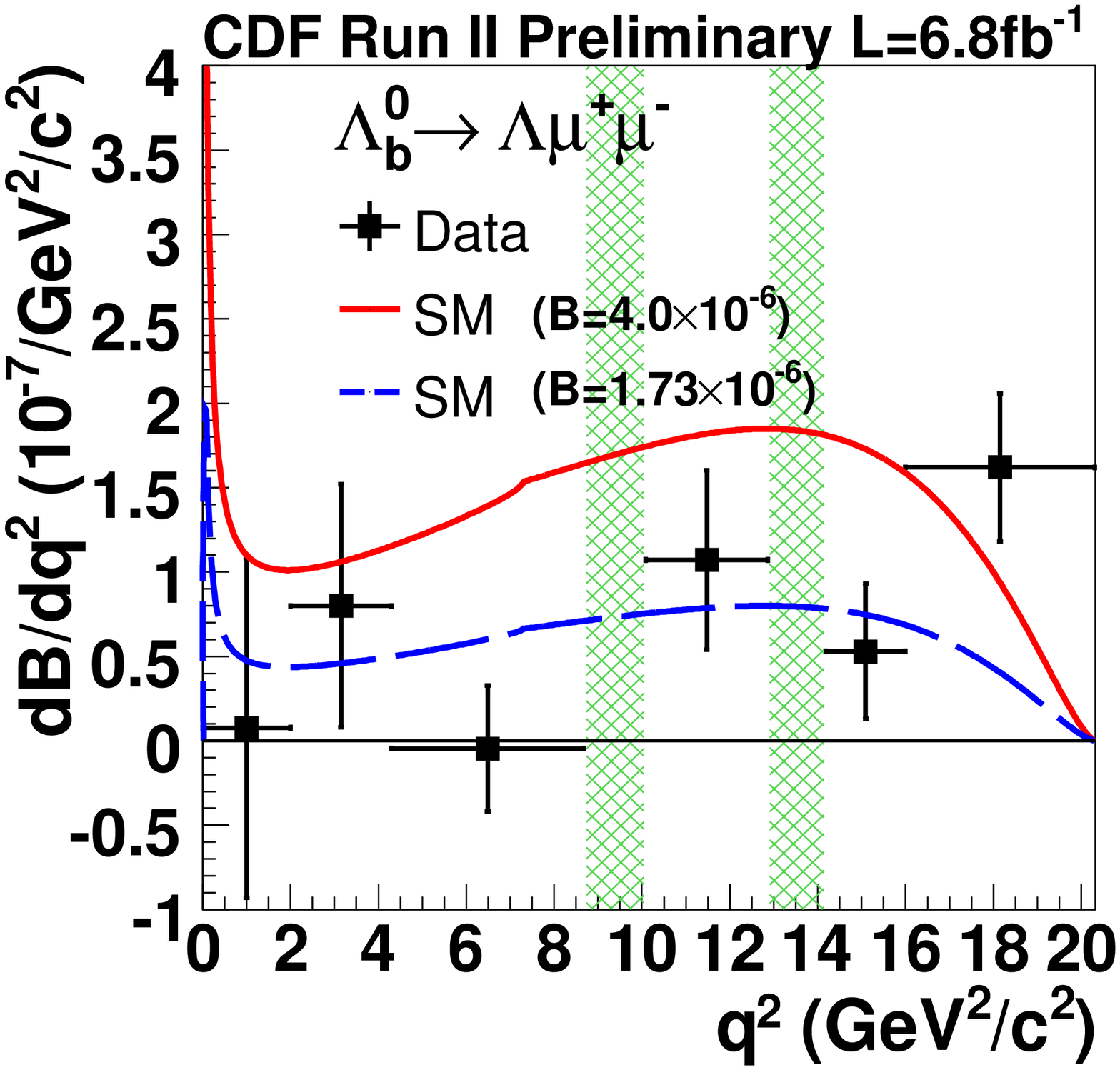}
}
\caption{
Differential branching ratios of 
(a) $B^+ \rightarrow K^+ \mu^+ \mu^- $, 
(b) $B^0 \rightarrow K^{*0} \mu^+ \mu^-$,
(c) $B^0_s \rightarrow \phi \mu^+ \mu^-$,
(d) $B^+ \rightarrow K^{*+} \mu^+ \mu^-$,
(e) $B^0 \rightarrow K_S\,\mu^+ \mu^-$, 
(f) $\Lambda_b^0 \rightarrow \Lambda \mu^+ \mu^-$.
The points are the fit result from data. The solid curves are the SM
expectation \cite{SMexp, SMexp2, SMexp3, SMexp4}. The dashed line in (f) is the SM prediction normalized to our total branching
ratio measurement. The hatched regions are the charmonium veto regions.}
\label{fig:bsmumudiff}
\end{figure}
%%%%%%%%%%%%%%%%%%%%%%%%%%%%%%%%%%%%%%%%%%%%%%%%%%%%%%%%%%%%%%%%%%%%%%%%%%%%%%

%Besides the total branching ratio, 
Rich information about the $b \rightarrow s \mu^+ \mu^-$
dynamics can be obtained by precise measurements of the differential branching ratio as a function of $q^2 = m_{\mu\mu}^2 c^2$
and the angular distributions of the decay products.
%%%%%%%%%%%%%%%%%%%%%%%%%%%%%%%%%%%%%%%%%%%%%%%%%%%%%%%%%%%%%%%%%%%%%%%%%%%%%%
\begin{figure}[h] 
  \centering
\subfigure[]{%\label{fig:afb}
  \includegraphics[width=5cm]{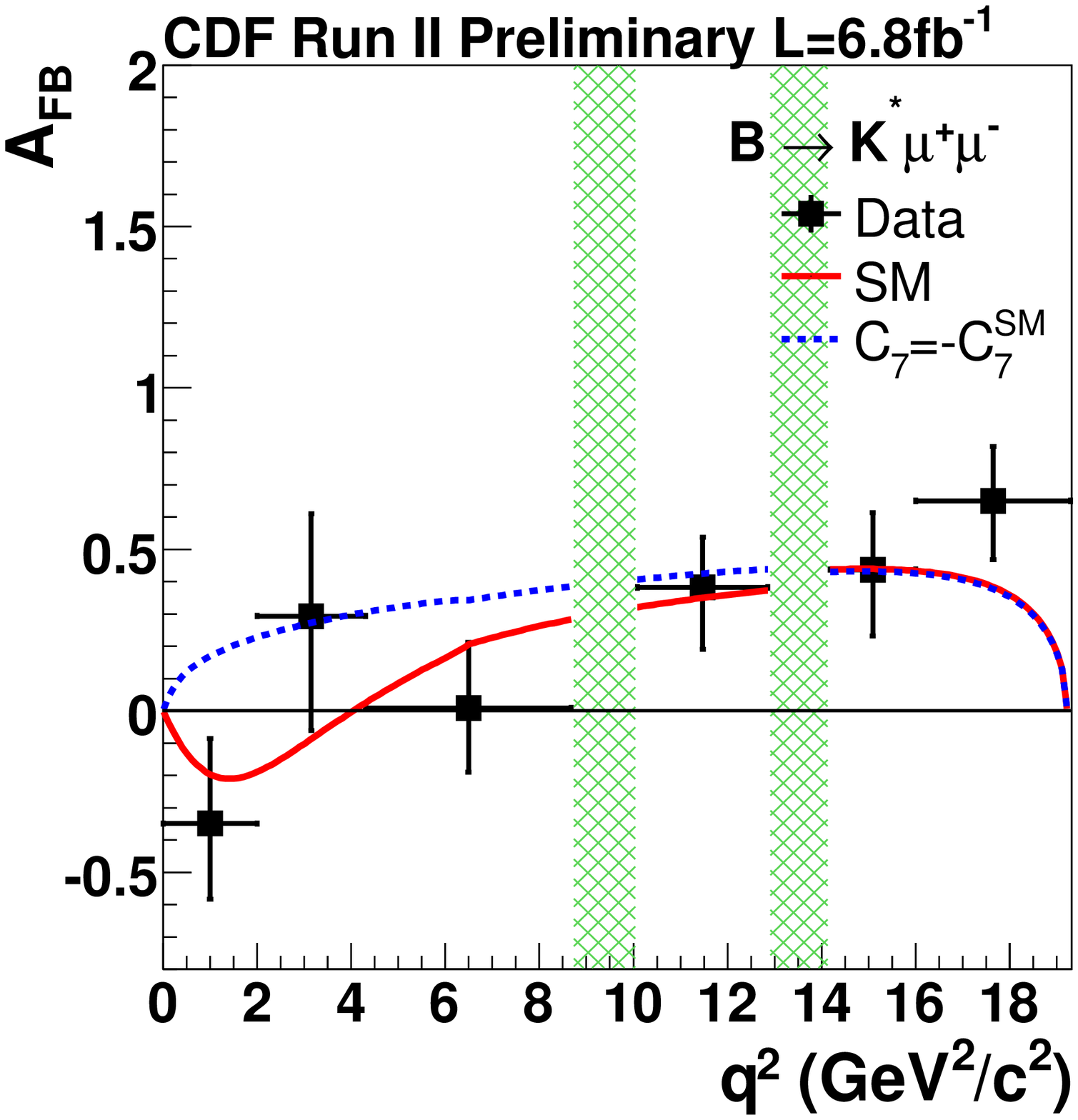}
}
\subfigure[]{%\label{fig:fl}
  \includegraphics[width=5cm]{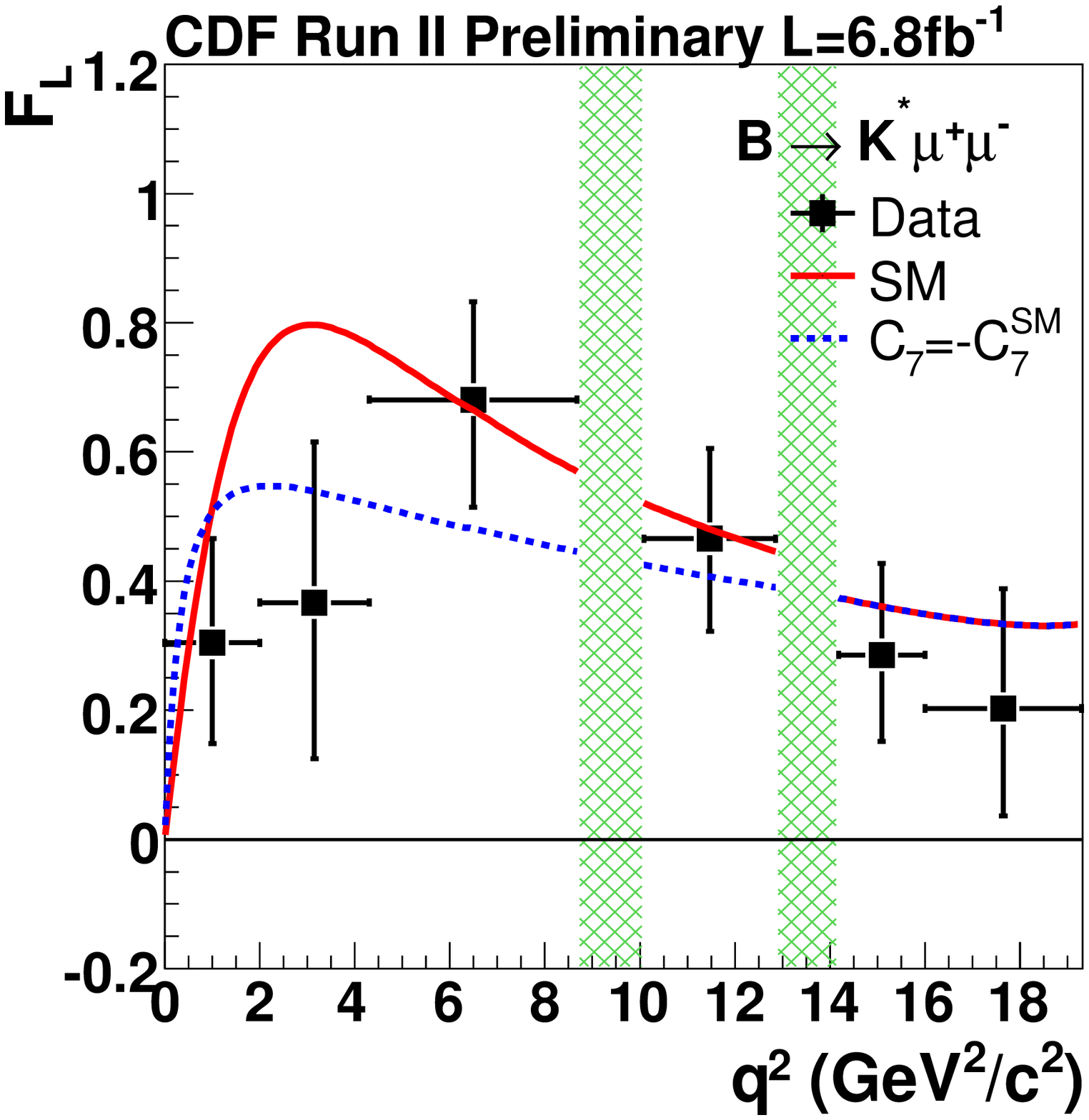}
}
\subfigure[]{%\label{fig:at2}
  \includegraphics[width=5cm]{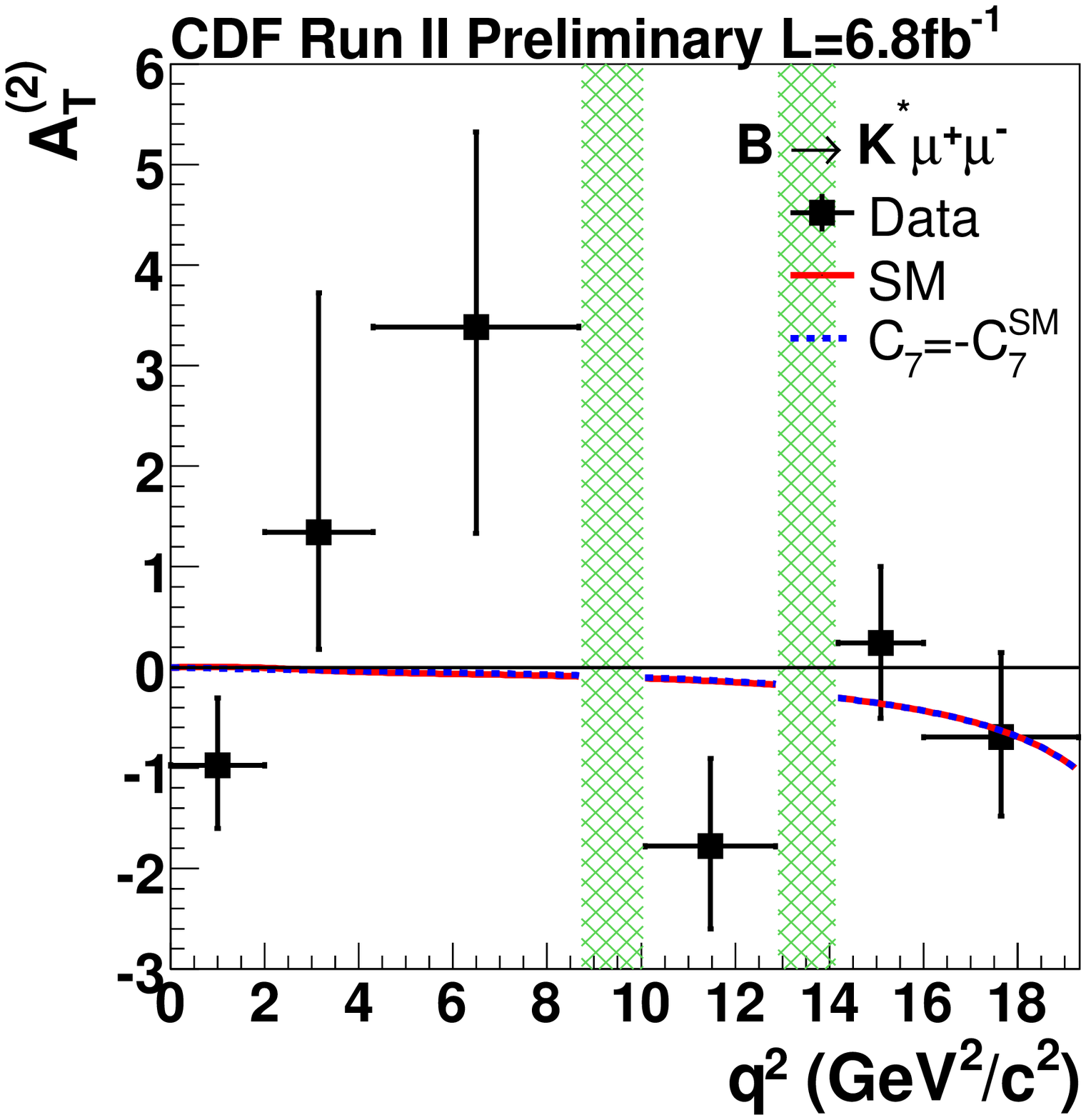}
}
\subfigure[]{%\label{fig:aim}
  \includegraphics[width=5cm]{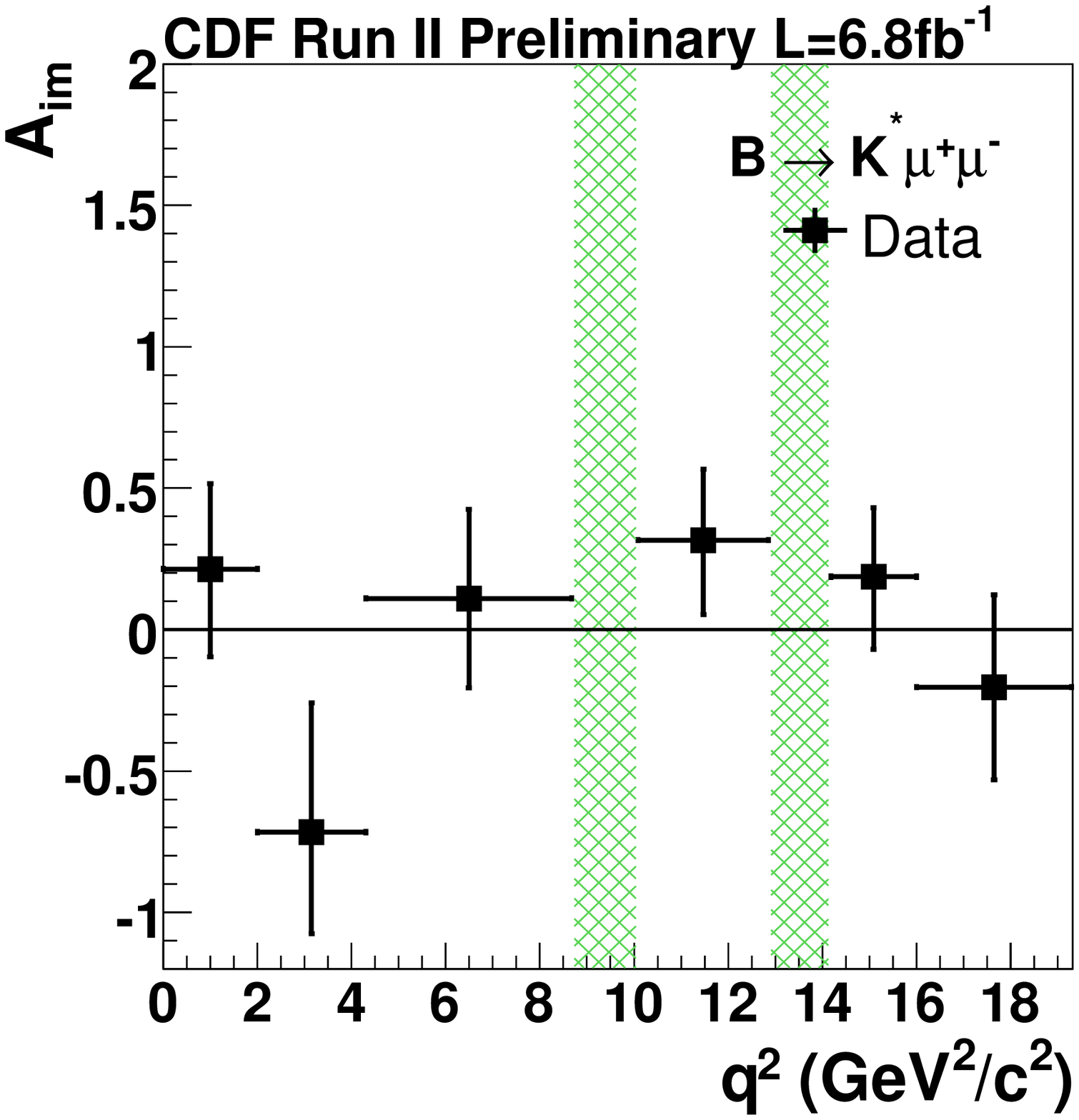}
}

\caption{Measurements of angular observables (a) $A_{FB}$, (b) $F_L$, (c) $A_T^{(2)}$, and (d) $A_{im}$ 
 as a function of dimuon mass squared $q^2$
in the combined decay mode $B \rightarrow K^{*} \mu^+ \mu^-$. 
The points are the fit results from data.
The solid and dotted curves represent expectations from the SM 
and a particular BSM scenario, respectively.}
\label{fig:angular}
\end{figure}
%%%%%%%%%%%%%%%%%%%%%%%%%%%%%%%%%%%%%%%%%%%%%%%%%%%%%%%%%%%%%%%%%%%%%%%%%%%%%%
The differential branching ratios with respect to $q^2$ have been measured by dividing 
the signal region  into six bins in $q^2$ 
and fitting the signal yield in each bin.
In each fit, the mean of the $H_b$ mass and the background slope were fixed to the
value from the global fit, so that only the signal fraction
was allowed to vary in the fit.
The results are shown in Fig.~\ref{fig:bsmumudiff}. 
For $B^0_s \rightarrow \phi \mu^+ \mu^-$ and $\Lambda_b^0 \rightarrow \Lambda \mu^+ \mu^-$ 
these are the first such measurements. At present no significant discrepancy from SM prediction is found.

%%%%%%%%%%%%%%%%%%%%%%%%%%%%%%%%%%%%%%%%%%%%%%%%%%%%%%%%%%%%%%%%%%%%%%%%%%%%%%%%%%%
The angular distributions 
of the combined  $B^0 \rightarrow K^{*0} \mu^+ \mu^-$
and $B^+ \rightarrow K^{*+} \mu^+ \mu^-$ decays have been measured and  
parametrized to four angular observables:
the muon forward-backward asymmetry $A_{FB}$, 
the $K^*$ longitudinal polarization fraction $F_L$,  
the transverse polarization asymmetry $A_T^{(2)}$,
the time-reversal-odd charge-and-parity asymmetry $A_{im}$, defined in \cite{ANG,ANG2}. 
$A_T^{(2)}$ and $A_{im}$  have been measured for the first time by CDF \cite{CDFbsmumuAfb}.
The results for these observables, shown in Fig.~\ref{fig:angular}, are among the most precise to date and consistent 
with SM predictions and other experiments, but
still statistically limited in providing stringent tests on
various BSM models.
%For instance predictions for Afb as a function of the dimuon invariant mass exist for several NP scenarios,  showing significant differences w.r.t. SM.
%BABAR BELLE CDF reported measurements of AFB larger than SM expecations.
%
%We have measured th BR total and differential of b hadrons decaying to a final state with two muons and a hadron where the hadron decays are listed in this table. 
%Moreover one can obtain sensitivity to BSM physics from precise measurement of angular distributions of the decay products.
\section{Conclusion}
We have summarized the recent updates on the searches of rare $b$-hadron decays at CDF.
The intriguing excess in $B_s^0 \rightarrow \mu^+ \mu^-$ reported in 2011
is confirmed with the full data set, though its significance is softened to the level of 2$\sigma$ over background. 
The measured $B_s^0 \rightarrow \mu^+ \mu^-$ branching ratio   is still compatible with 
the SM expectation and recent combined results from LHC experiments.

The dynamics of several rare decays mediated by the FCNC process $b \rightarrow s \mu^- \mu^-$ 
has been studied in detail extending the reach to new angular observables. 
The $\Lambda_b^0 \rightarrow \Lambda \mu^+ \mu^-$ decay has been observed for the first time. 
Analyses of the  $b \rightarrow s \mu^- \mu^-$  decays 
are still in progress and may yield interesting results in the
near future.

\end{document}